# Interacting nanopatterned pins: barriers towards creating interstitial vortices

Gorky Shaw, Shyam Mohan, Jaivardhan Sinha and S. S. Banerjee[a]

Department of Physics, Indian Institute of Technology, Kanpur-208016, U. P., India

We show that by nano-patterning a superconductor ($NbSe_2$ single crystal) with an array of blind holes produces significant magnetic field sweep rate dependent metastable magnetization response. Our results are explained on the basis of a unique collective action of the blind holes pins which creates a barrier against vortex redistribution inside the sample. We propose that this barrier leads to a phase separation creating distinct population of vortices viz., those pinned on blind holes and those confined in the interstitials between the holes.

[a] email: satyajit@iitk.ac.in

Competition and interplay between disorder, thermal fluctuations and interaction between vortices leads to a variety of phases. For example, under the influence of enhanced thermal fluctuations there is thermal melting of the ideal Abrikosov vortex lattice phase into the vortex liquid phase[1,2]. Under the influence of disorder the ideal hexagonally ordered vortex lattice loses long range order and exhibits a diverse variety of glassy phases[1]. In recent times a new focus on metastable novel phases of vortices under extreme confined conditions in mesoscopic superconductors[3]. Extreme confinement is generated by physically patterning the dimensions of the superconductor to be comparable to the superconducting coherence length ($\xi$) or the penetration depth ($\lambda$). In such mesoscopic superconductors exotic states viz., giant multiquanta fluxons are known to exist[3], apart from the appearance of significant metastable vortex configurations. Vortices also have the possibility of being confined in the interstitial spaces between columnar defects (CDs)[4], produced with heavy ion irradiation of samples. The disadvantage with CDs is that its location is random and not always well controlled. Controlled well ordered extended defects (pinning centers) can be produced by nanoscale patterning of the superconducting material. A lot of work[5,6,7,8,9,10,11,12,13,14,15] has focused on the nature and strength of pinning generated by these patterned pinning centers. The most often studied nanoscale patterned pinning center is the well ordered array of antidots[5-15], which are cylindrical holes with a diameter of $O(nm)$ carved out through the entire thickness of a superconductor. It is found that when the periodically spaced vortices in the superconductor (intervortex spacing, $a_0 \propto \sqrt{\phi_0/B}$ where $\phi_0 = 2.07 \times 10^{-7}$ G-cm$^2$ is the magnetic flux quantum of a single vortex and $B$ is the magnetic field) matches with the periodic array of antidot holes with a period $d$, there is a large enhancement in pinning. In this paper instead of focusing on the often investigated pinning properties of the nanopatterned holes, we are interested in investigating certain open issues viz., how does the presence of these nano-patterned holes affect vortices around the holes. Do the vortices in the interstitial spaces between the patterned holes have properties similar to that in sample without these holes? Does it cost extra energy to get interstitial vortices in between the holes? Can these patterned pinning centers lead to confinement of vortices in the interstitial spaces between the holes? Similar to antidots is a related structure called blind holes[9,15] which are cylindrical holes with diameter of $O(nm)$, closed at one end by the material in which they are patterned. However, compared to antidots, the less investigated blind holes are considered to be weaker pins[15], as a result we had expected that with blind hole patterning the pinning in the sample will be only weakly perturbed. However, we show in this paper that in pure single crystals patterned with relatively small number of blind holes, the bulk properties of the vortex state in the sample get substantially modified. We find the appearance of significant metastable magnetization response in the patterned sample. We discuss our results in the light of a unique collective action of nanopatterned blind holes within the sample.

Unlike most other previous investigations on nanopatterned thin film[9,15,16] we study much thicker structures. Compared to thin films, single crystals of $NbSe_2$ have lower[17] $J_c/J_0 \sim 10^{-6}$ (where $J_0$ is the depairing current density) which implies much weaker intrinsic sample pinning. Two NbSe$_2$ crystals belonging to a batch[18] of high quality single crystals (1.5 *mm* x 1 *mm* x 30 *μm*, $T_c$ = 7 *K*) were investigated. One of the crystals was milled with a focused gallium (Ga) ion beam (diameter ~ 7 *nm*) using the Focused Ion Beam (FIB) machine (dual beam FEI make Nova 600 NanoLab machine at IIT Kanpur, India) to produce a hexagonal array of blind holes, each with a diameter of 170 *nm* with a mean spacing of 350 *nm*, covering a total area of 180 *μm²* away from the sample edges (cf. Fig.1). For anisotropic NbSe$_2$ the typical values of the intrinsic parameters for the ab – plane are, $\xi_{ab}$ ~ 7.7 *nm* and penetration depth $\lambda_{ab}$ ~ 120 *nm*[19]. Note that only ~ 0.02% of the total sample area was patterned. Inset, fig.1(b), shows a Scanning Electron Microscopic (SEM) image of the magnified portion of the entire patterned area (cf. Fig.1 main panel).

Bulk dc magnetization hysteresis *M(H)* loops of patterned and unpatterned NbSe$_2$ samples were measured at different temperatures (5.0 *K*, 5.5 *K*, 6 *K* and 6.7 *K*, all data not shown) with a Quantum Design Superconducting Quantum Interference Device (SQUID) magnetometer (using the RSO option to reduce field inhomogenity artifacts, cf. http://www.qdusa.com/resources/pdf/mpmsappnotes/1014-820.pdf). Figure 2 shows nearly identical *M(H)* loops of the patterned and unpatterned NbSe$_2$ crystals. The inset of fig.2 shows the ratio of the widths of *M(H)* loops for the patterned and unpatterned samples i.e., $\Delta M_{patt}/\Delta M_{unpatt}$ as a function of *H* is ~ 1 - 1.5. As the width of the magnetization hysteresis ($\Delta M$) is a direct measure of the pinning strength in the superconductor[20], it appears from above ratio that pinning in the sample is only weakly perturbed by the blind holes. Such a feature is not unexpected as a large area of the sample was unpatterned and therefore the *M(H)* appears to be dominated by the weakly pinned vortices in the unpatterned regions of the sample.

Using a Vibrating Sample Magnetometer (VSM) (Oxford make, Model 3001), *M(H)* loops were measured for different magnetic field (*H*) sweep rates of nearly 0.05, 0.1, 0.2, and 0.3 *T/min*. In Fig.3, we compare the *M(H)* loops for the patterned sample recorded with a sweep rate of nearly *0 T/min* (on SQUID) and at a high sweep rate of 0.2 *T/min* (on VSM). We observe that, for same *H*, the *M* values differ with the sweep rate, indicating a significant metastable magnetization response. Inset of Fig.3 compares $\Delta M_{patt} / \Delta M_{unpatt}$ for two different sweep rates. From this it is clear that, for a range of magnetic field values (upto 0.4 *Tesla*), the ratio $\Delta M_{patt} / \Delta M_{unpatt}$ at field sweep rate of 0.2 *T/min* is significantly higher, by about 5 – 6 times than that for 0 *T/min*. The sweep rate dependence of the width of the *M(H)* loop shows that the patterned sample seems to explore a stronger pinning landscape at larger sweep rates ($\geq$ 0.2 *T/min*).

Figure 4 shows the manifestation of the strong pinning landscape through the behavior of the irreversibility field, $H_{irr}(T)$ at different sweep rates. Above $H_{irr}(T)$ pinning effects diminish thereby leading to *M(H)* becoming path independent viz., $M_{for}$ and $M_{rev}$ coincide. It is clear from fig.3 that in the patterned sample at 5 *K* at 0.2

*T/min*, the M-H loop has a higher irreversibility field $H_{irr}(T)$ as compared to that for a stabilized field measurement. Comparing $H_{irr}(T)$ for the patterned sample obtained from measurements at stabilized field mode ($H_{irr,0T/min}$) and at a field sweep rate of 0.2 *T/min* ($H_{irr,0.2T/min}$) (cf. fig.4), we find that $H_{irr}(T)$ shifts upwards with higher magnetic field sweep rates, reaffirming the results from fig.3. It is to be noted that, commensurate with the observations in fig.2, $H_{irr,0T/min}$ is identical for both patterned (solid (blue) square) and unpatterned samples (open (blue) square).

The sweeping of magnetic field drives the vortices as they redistribute themselves inside the superconductor. The driven vortices generate an electric field, $E=u\bullet B$, where *u* is the drift velocity of the vortices. An approximate procedure used to construct the electric field from the magnetization data[21] is $E \propto dM/dt$, or $E \propto A\times(dM/dH)/(dH/dt)$; where ($dH/dt$) is the magnetic field sweep rate and *A* is the sample area). The width of the hysteresis loop ($\Delta M$) provides information about the shielding current[20], *J* induced in the superconductor as *H* is swept (($\Delta M/w) \propto J$, *w*: mean width of the sample). The *E - J* curves so deduced from the sweep dependent *M(H)* loops are analyzed with the relationship invoked for collective vortex creep in vortex glasses[22,23] viz., $E \propto \exp\left[-\left(U_c/kT\right)\left(J_c/J\right)^\mu\right]$; where $U_c$ is the depth of the potential well at $J = J_c$, and *T* is temperature. Inset of fig. 4 compares the *E - J* curve determined from the $M_{for}(H)$ curve recorded with 0.1 *T/min* and 0.2 *T/min*. The dashed (red) and solid (blue) lines correspond to the *E(J)* fitted to the data using the above expression with a fixed $\mu = 3$ and $\mu = 1$ respectively, with $U_c$ and $J_c$ as variable fitting parameters. By extrapolating the fitted lines, we find two intercepts $J_{c,in}$ and $J_{c,bh}$ on the $J$ axis. In general, notice that the $E$ - $J$ for low field sweep rates (of about 0.1 *T/min* and below) has the lower $J_{c,in}$ intercept. Theory predicts that for weak collective pinned vortices in the sample $\mu \sim 1$ to $1.5$ (see M. V. Feigel'man Ref. 23). The likely source of the lower intercept ($(\Delta M/w) \propto$) $J_{c,in} \approx 0.94 \times 10^3$ A/m$^2$ is the weak intrinsic pinning centers (with $\mu = 1$) in the unpatterned regions. We believe that the higher $J_{c,bh} \approx 1.98 \times 10^3$ A/m$^2$ intercept (with $\mu > 1$) at high sweep rate, corresponds to the enhanced depinning threshold from the stronger blind hole pinning centers. While inset of fig.4 gives an approximate comparison of the two $J_c$'s in our sample, a comparable independent estimate of the same is obtained from the ratio $\Delta M_{patt}/\Delta M_{unpatt}$ in Fig.3. (Note this increased $J_c$ is obtained by patterning only 0.02% area of the crystal).

As the thickness of the superconductor below each blind hole cylinder is less than outside it therefore, the blind holes are the energetically more favorable sites for creating and pinning vortices than anywhere else. As the spacing ($d$) between the blind holes is 350 nm, therefore at *195* G (where $a_0 \sim d$), a good fraction of the blind hole sites are occupied. We now investigate the ease with which extra vortices can be placed in the interstitial spaces between the blind holes, viz., the interstitial vortices. Recently for antidots in a superconductor placed in a magnetic field, based on Ginzburg - Landau equations, the screening current distribution was shown[24] to be symmetric around with hole and decays as $exp(-r/\lambda)$ as one moves radially ($r$) outwards from the antidot. Shown in the lower inset of Fig.4 is the calculated screening current density ($J$) distribution around four blind holes in a 2-dimensional planar superconductor. The four blind holes are located at the corners of the dotted black square. The $J$ distribution around each hole is modeled[24] as $\vec{J}(r,\theta) = J_{c,bh}\exp(-r/\lambda)\hat{\theta}$. In our calculation, the spacing ($d$) between blind holes was chosen to be close to our patterned sample viz., *d = 350 nm ~ 3λ*. Using the above form for $J$, we calculate the net screening current density in each pixel of the planar superconductor around the blind holes. The white in the image corresponds to a value of $J = J_{c,bh}$, while the black corresponds to $0.125 J_{c,bh}$. Upto the depth of the blind holes (~ *1* μm in our case), due to similarity in the structure, the $J$ distribution calculated around the blind hole is similar to that in antidots. At a depth below the termination of the blind hole, the calculated $J$ circulating around the vortices pinned on the blind holes have a form similar to that at the surface. It is clear from the lower inset of Fig.4 that around the blind holes with $d$ close to $\lambda$, the screening currents around each blind hole is distorted by the presence of neighboring blind holes. As a result due to the overlapping screening current from neighboring blind holes, appreciable screening current (~ *0.3$J_{c,bh}$*, dark brown shade) circulates in a closed path around the four blind holes. Thus for $d \leq \lambda$, there appears to be an interaction between the physically separated blind holes via the screening currents, due to which they can act collectively.

Lower inset of Fig.4 shows pockets (nearly circular region in black) in the center of the dotted black square as well as along the edges which don't possess any appreciably large screening currents. These black pockets are regions around the blind hole array where interstitial vortices can be trapped between the blind holes. However to bring vortices at the site of these pockets a barrier due to the screening currents (brown shade) circulating around the blind holes array must be overcome. The sweeping magnetic field helps to drive vortices over the barrier around the patterned area. Once the barrier is overcome the vortices can occupy the interstitial pockets between the blind holes as they possess low screening current density. Based on the above calculation, we expect that along the edges of the *180 µm$^2$* patterned area screening currents circulate creating a barrier against vortices occupying the interstitial spaces between the blind holes. However once the barrier is overcome with large field sweep, these vortices get trapped and confined at the interstitial spaces. From earlier studies[4] of interstitial vortices in samples with dilute density of columnar defects it is known that due to the strong confinement thermal fluctuations of the interstitial vortices are suppressed and hence the $H_{irr}$(T) line is shifted upwards. A similar effect is observed in our sample with blind pinning array (cf. $H_{irr,0.2T/min}$(T) line in Fig.4).

To summarize, the distinct behavior of nanopatterned blind holes arise due to their unique ability to interact with each other (via a screening current) when d ~ λ. The collective action of blind pins results in a barrier which phase separates distinct population of vortices, viz., vortices pinned on blind holes, vortices pinned in weak intrinsic defects in the unpatterned regions in the sample and those pinned in the interstitial regions between the blind holes. Metastability arises due to certain configurations of these three distinct populations getting arrested with reasonably large relaxation times. We hope this work will pave the way to understand the properties of the heterogeneous phase of vortices in the samples with nanopatterned blind holes.

SSB would like to acknowledge the funding support from Department of Science and Technology (DST), India and CSIR India. SSB would also like to thank Prof. P L Paulose and Prof. A. K. Grover for making facilities available at TIFR, Mumbai.

**Figure Captions:**

**Fig.1:** Scanning electron microscope (SEM) image of the region of the NbSe$_2$ crystal patterned with a hexagonal array of blind holes. Inset (a), SEM image of the larger area of the sample. The red-cross indicates the micron sized patterned area. Inset (b), magnified image of the patterned region showing uniform hexagonal pattern of blind holes (cf. text for details)

**Fig.2**: Isothermal bulk dc magnetization hysteresis *M(H)* loops of the patterned and unpatterned NbSe$_2$ samples (denoted by circles and triangles respectively) measured at 6 *K* using SQUID magnetometer. Inset, ratio of the widths of *M*(*H*) loop, for the patterned and unpatterned samples i.e., $\Delta M_{patt} / \Delta M_{unpatt}$ as a function of *H*.

**Fig.3:** The main panel compares at 5 *K* the *M(H)* loops for the patterned sample recorded with nearly zero and 0.2 *T/min* (red curve, open circles) field sweep rate. The inset shows $\Delta M_{patt}/\Delta M_{unpatt}$ as a function of *H* for 0 *T/min* (blue curve, open circles) and 0.2 *T/min* (red curve, closed circles) (cf. text for details).

**Fig.4.** $H_{irr}(T)$ line for the patterned sample for 0.2 *T/min* (solid (red) line) ($H_{irr,0.2T/min}$) and for stabilized field measurement (dotted (blue) line) ($H_{irr,0T/min}$). Solid (blue) squares indicate $H_{irr,0\,T/min}(T)$ for the patterned sample and open (blue) squares are for the unpatterned sample. Inset, [A × (dM$_{for}$/dH)/dH/dt] versus (ΔM/d) for 0.1 *T/min* and 0.2 *T/min* field sweep rates for the patterned sample. (cf. text for details). Lower inset show the calculated distribution of screening currents around four blind hole array. The color bar represents the magnitude of the current distribution in units of $J_{c,bh}$ (refer to text for details).

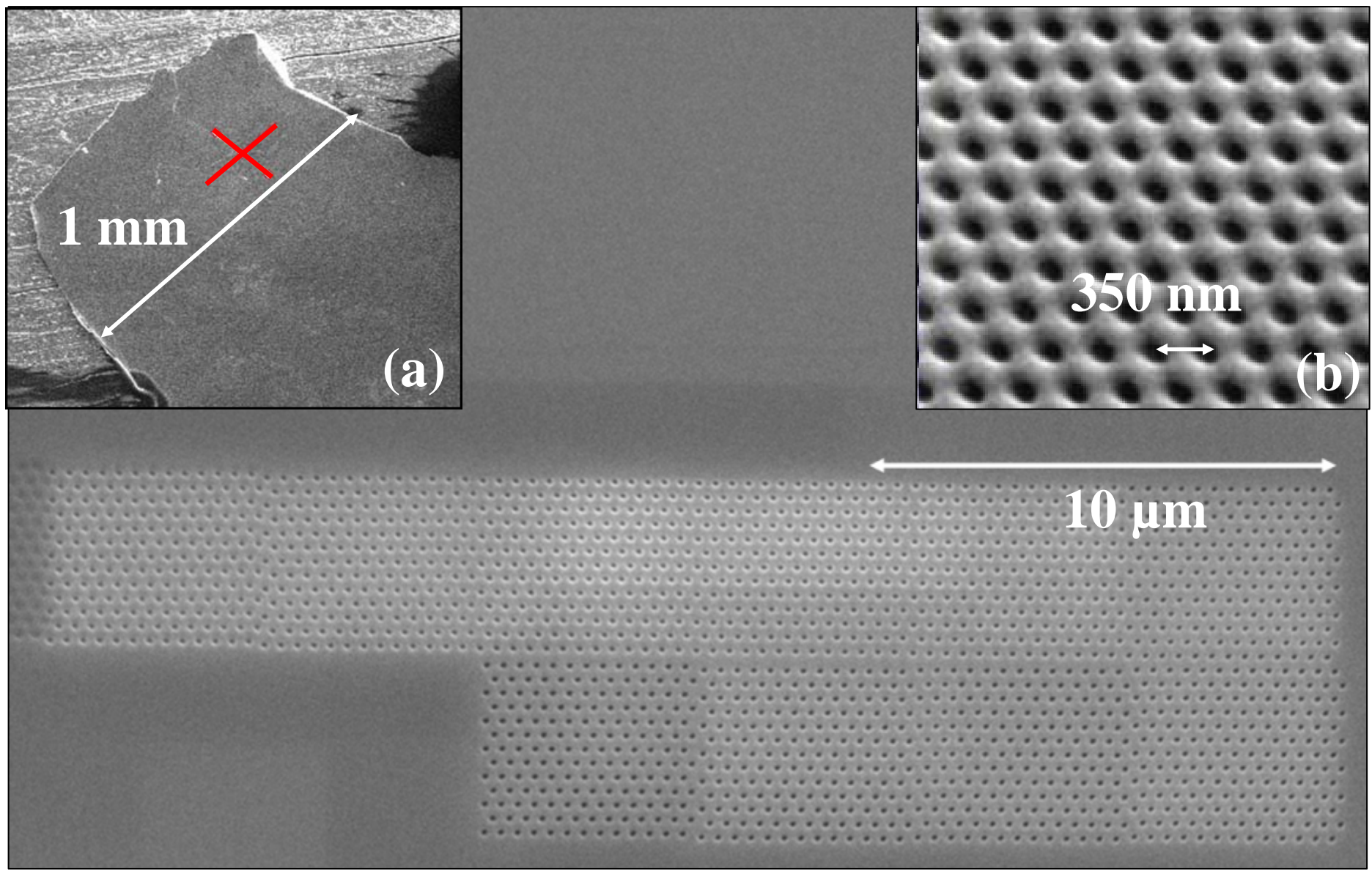

Fig 1

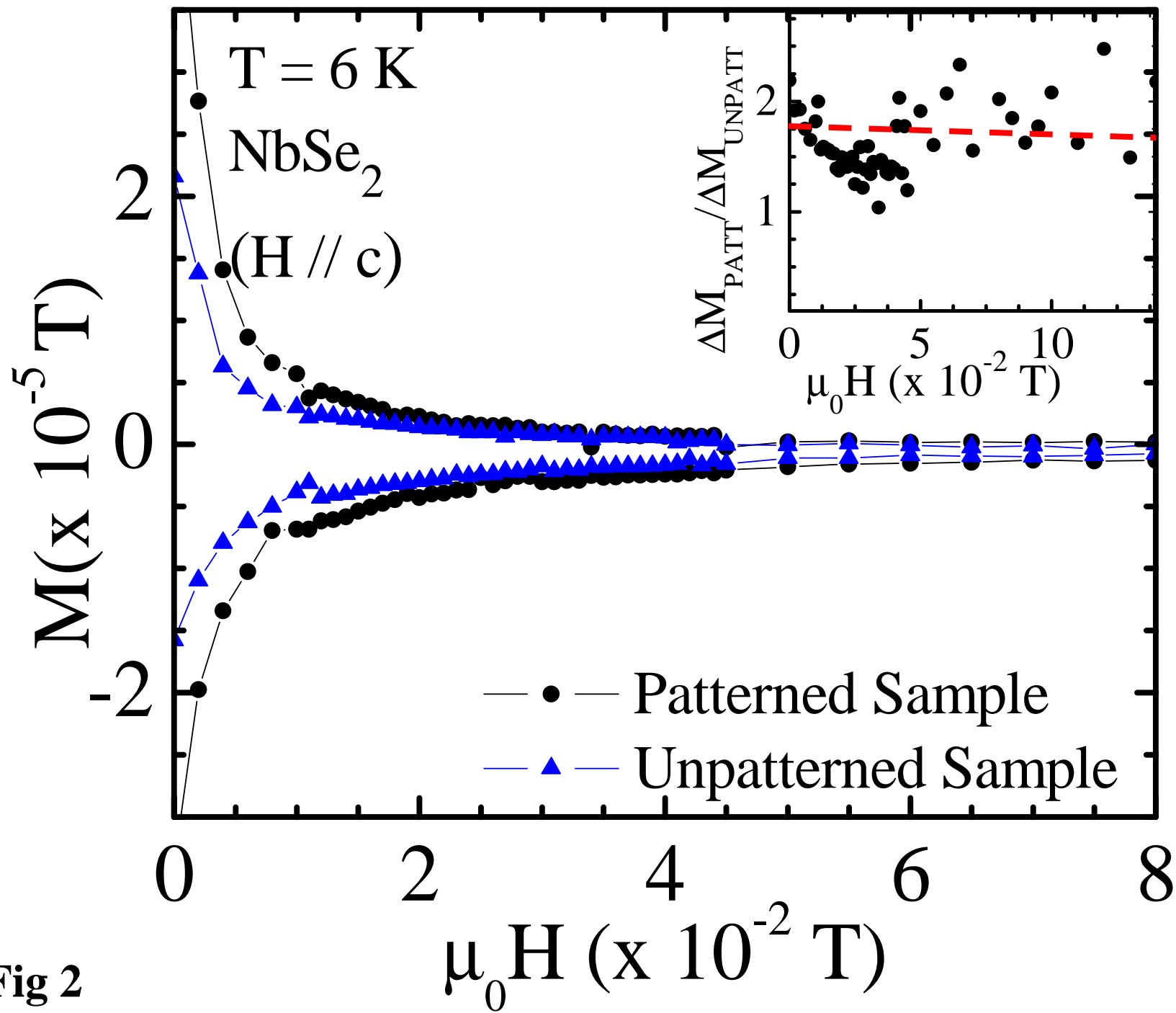

**Fig 2**

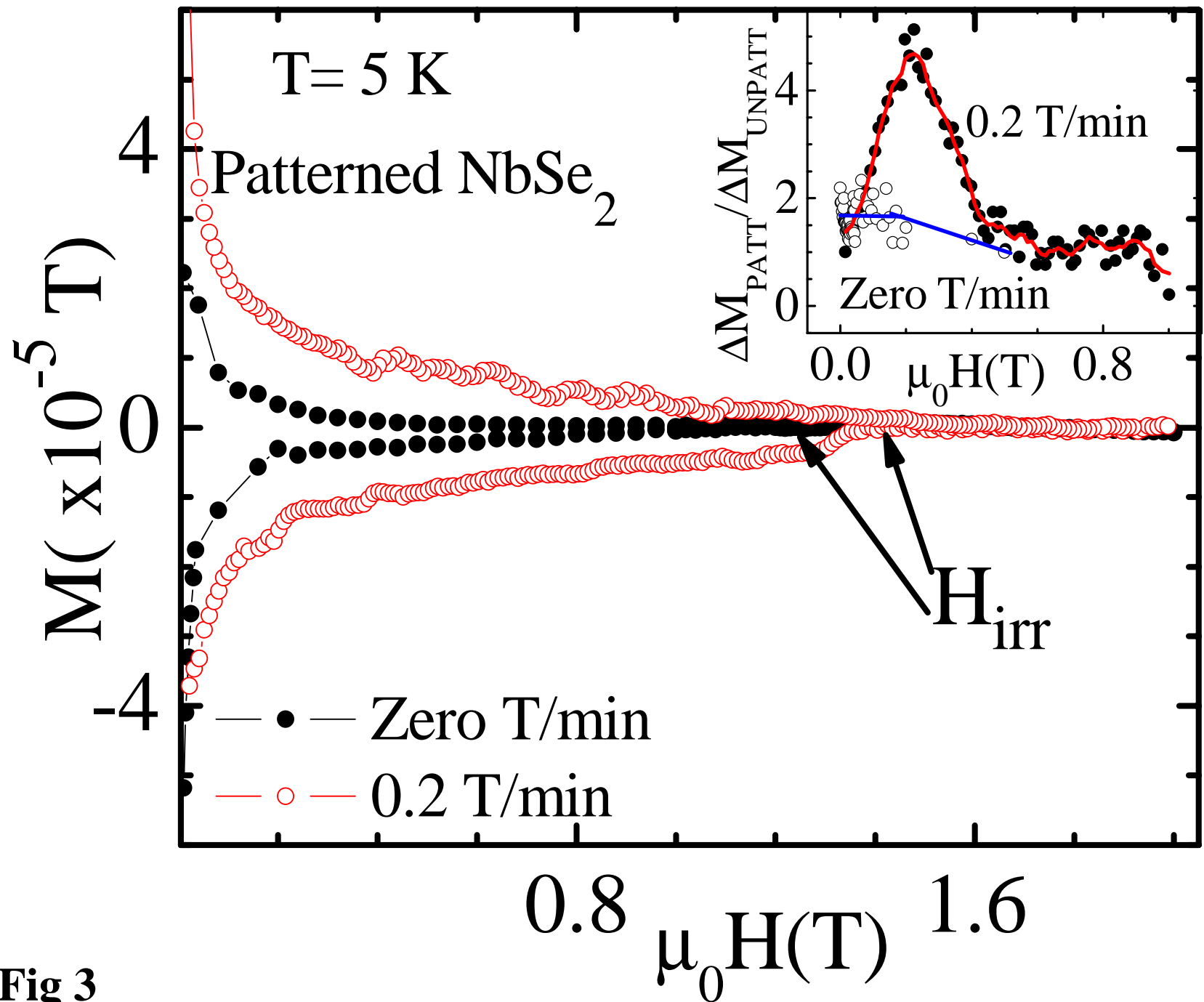

Fig 3

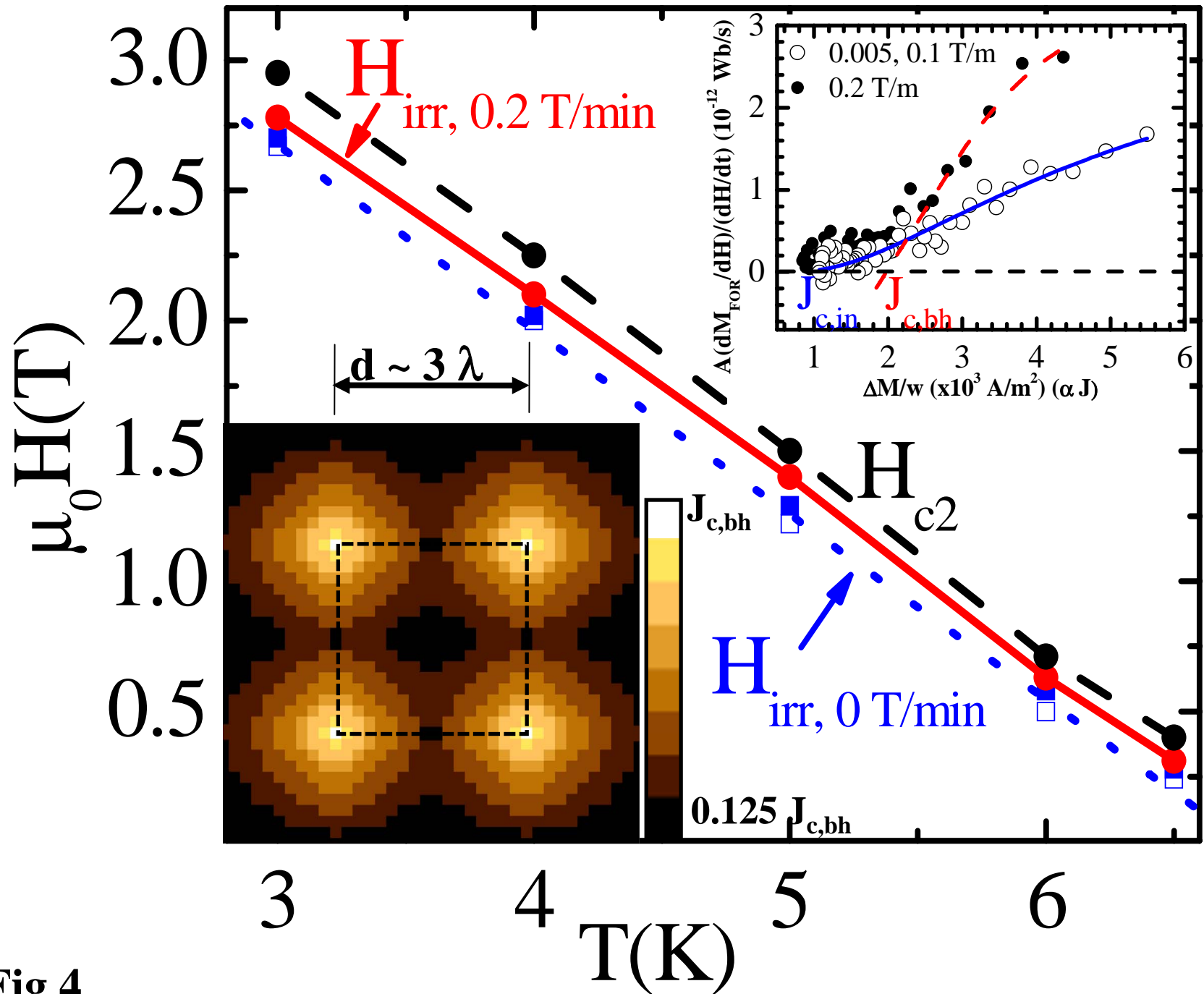

Fig 4